\documentclass[a4paper,11pt]{article}
\usepackage{graphicx}
\usepackage{dcolumn}
\usepackage{bm}
\usepackage{amsmath,amsfonts,amssymb}
\usepackage{subcaption}
\usepackage{setspace}
\usepackage{color}
\usepackage{tikz}
\usepackage{float}
\usepackage{hyperref}
\usepackage{parskip}
\usepackage[normalem]{ulem}
\title{\textbf{When Black Holes Relax in a Cold Bath:\\
Evolving Page Curves for Black Holes Coupled to Cooling Baths}}

\author{
\textbf{Waheed A. Dar}\thanks{Email: \href{mailto:waheed.dar1729@gmail.com}{waheed.dar1729@gmail.com}} \\
\small Department of Physics, National Institute of Technology Srinagar,\\
\small Kashmir-190006, India
\and
\textbf{Nirmalya Kajuri}\thanks{Email: \href{mailto:nirmalya@iitmandi.ac.in}{nirmalya@iitmandi.ac.in}} \\
\small School of Physical Sciences, Indian Institute of Technology Mandi,\\
\small Himachal Pradesh, India
\and
\textbf{Rinkesh Panigrahi}\thanks{Email: \href{mailto:d23189@students.iitmandi.ac.in}{d23189@students.iitmandi.ac.in}} \\
\small School of Physical Sciences, Indian Institute of Technology Mandi,\\
\small Himachal Pradesh, India
}

\date{} 

\begin{document}

\maketitle

\begin{abstract}
We consider scenarios where Jackiw-Teitelboim black holes are attached to baths whose temperatures can be manipulated externally. We consider the bath to be cooled continuously and numerically investigate the subsequent evolution of the generalized entropies of the black hole. Page curves corresponding to several different cooling profiles of the bath are obtained.
\end{abstract}

\section{Introduction}

The evaporation of black holes has been a topic of fascination for physicists for many years. According to the principle of unitarity, the entanglement entropy of radiation should initially increase but eventually decrease during the evaporation of a black hole, following what is known as the Page curve \cite{Page:1993df}. Obtaining a Page curve for black hole evaporation remained an open challenge for a long time. Only recently was it shown that for certain black holes, the entanglement entropy during evaporation indeed follows the Page curve \cite{Almheiri_2019, Penington:2019npb, Almheiri:2019hni}.

In \cite{Almheiri:2019hni}, for example, the set-up considered was a two-dimensional black hole in Jackiw-Teitelboim gravity attached to a bath. By computing the quantum extremal surfaces via the generalized entropy formula of Engelhardt and Wall \cite{Engelhardt:2014gca}, the Page curve was recovered. Since then, there has been significant progress in understanding the results and extending them to various black holes \cite{Penington:2019kki, Almheiri:2019qdq, Almheiri:2020cfm,Almheiri:2019yqk, Marolf:2020xie, Almheiri:2019psy, Chen:2020uac, Hashimoto:2020cas, Hartman:2020khs, Gautason:2020tmk, Balasubramanian:2020xqf,  Krishnan:2020fer, Krishnan:2020oun, Ling:2020laa, Ahn:2021chg}.  

Most existing derivations of the Page curve in JT gravity consider stationary or quasi-stationary situations, in which the bath temperature is fixed and the black hole settles into equilibrium. In such cases, the quantum extremal surface (QES) either remains fixed or evolves adiabatically. However, the evaporation of a black hole can be influenced by external conditions such as changes in ambient temperature or energy exchange with other systems. For instance, primordial black holes in the early universe evaporate in an environment with rapidly changing ambient temperature. 

Once the bath temperature becomes explicitly time-dependent, the geometry is no longer stationary and the QES must move dynamically. The Page time may shift, the entropy production rate of Hawking radiation changes continuously, and the generalized entropy must interpolate between distinct equilibrium configurations.
A natural question to ask is: how does changing the external conditions in a time-dependent manner affect the Page curve? In other words, how robust is the Page curve under time-dependent externalities?

In this paper, we address this question in the same setup considered in \cite{Almheiri:2019hni}: a 2D black hole coupled to a reservoir. In our case, the reservoir is taken to be a \textit{cooling} bath: one with controllable, time-dependent temperature. This provides a simple toy model for non-equilibrium evaporation. our goal is to understand how the Page curve evolves when the black hole is driven away from equilibrium by a time-dependent external environment. 

Throughout this work we remain in the semiclassical regime, where backreaction is small and the generalized entropy formula reliably captures the Page curve. Our setup therefore provides a controlled framework to study non-equilibrium evaporation while remaining within the semiclassical domain where the island formula applies.

The results from this model carry over beyond the case of a cooling bath. They apply to all processes where the black hole loses (or gains) mass in a time-dependent manner. Examples include the injection of negative mass shells into the black hole \cite{Cadoni:2021ypx} and partially reflecting boundary conditions with time-dependent reflectivity. As we will show, all such setups for a black hole out of equilibrium with the bath are dependent on a single parameter, and hence mathematically equivalent.  

 In JT gravity, the black hole geometry is described by boundary parameterization $f(t)$. As the temperature of the bath changes, it leads to a change in $f(t)$. This in turn shifts the location of the quantum extremal surface (QES), thereby modifying the Page curve. This is the modification that we aim to study in this paper. There are a few special situations where analytic control is possible, such as instantaneous temperature jumps sourced by shock waves from the bath to the black hole \cite{Hollowood:2020cou}. Outside of these cases, the evolution of $f(t)$, and therefore the Page curve, can not be computed analytically. 
 
 We therefore adopt a  numerical approach that allows for arbitrary profiles beyond analytic solvability. We consider scenarios where the bath temperature is allowed to change \textit{continuously} with time and compute the modified Page curves numerically for various cooling profiles. In the limiting case of shock waves, we demonstrate that our results are in agreement with those of \cite{Hollowood:2020cou}, where they were computed analytically.

In more detail, we consider three different cooling profiles. They are: an instantaneous temperature drop (for the bath), linear drop in energy, and the temperature following Newton's law of cooling. These profiles are chosen to model environments where a black hole exchanges energy with an external system whose properties evolve in a controllable way. Our only assumption is that the process is slow enough to remain within the semiclassical regime (back-reaction remains negligible).

For each case, we track the subsequent evolution of the boundary parameterization numerically utilizing the RK4 method, starting from the parameterization corresponding to a black hole at temperature $\beta$ and reaching a temperature $\tilde\beta$. We compute the positions of the quantum extremal surfaces and the corresponding entropies at each time step during the relaxation to a new temperature. The results are the Page curves of evaporating black holes in their non-equilibrium phases for each of the three cooling profiles.

Note that while we have restricted ourselves to these profiles, the numerical method we use can be utilized for any general case as long as it stays in the semiclassical regime.

To summarize, our main new contribution is extending the computation of the Page curve to cases where the black hole is interacting with a controlled, time-dependent environment (which changes the boundary parameterization in a continuous manner) utilizing numerical methods.

The structure and outline of the paper are set out as follows. In Section II,  we start with preliminaries and provide the setup. Section III presents the computations for the instantaneous drop in temperature. In Section IV, we generalize to more general cooling profiles. We conclude with a brief summary of the new results and future directions.

\section{Preliminaries}
\subsection{Jackiw-Teitelboim Gravity}
We start by reviewing the basic aspects of Jackiw-Teitelboim (JT) gravity \cite{Jackiw:1984je,Teitelboim:1983ux} and nearly AdS$_2$ holography \cite{Maldacena:2016hyu,Almheiri:2014cka,Jensen:2016pah,Maldacena:2016upp,Engelsoy:2016xyb}. We refer the reader to \cite{Mertens:2022irh} for a recent detailed review. 

The action of JT gravity minimally coupled to matter is given by:
\begin{eqnarray}
  I[g,\Phi]=I_{\text{grav}}[g, \Phi]  + I_{\text{matter}}
  \label{1}
\end{eqnarray}
Here, $\Phi$ denotes the dilaton field. 
The gravitational action is given by:
\begin{align} \notag
I_{\text{grav}}[g, \Phi]=\frac{-1}{16 \pi G_{N}} \int_{\mathcal{M}} \sqrt{g} & \Phi \left(  R + 2 \right) - \frac{1}{8 \pi G_{N}} \int_{\partial \mathcal{M}} \sqrt{h}  \Phi (K - 1) 
\\ &-\frac{\Phi_0}{16 G_N \pi }\left(\int_{\mathcal{M}} \sqrt{g}R+2 \int_{\partial \mathcal{M}} \sqrt{h} K\right)
\end{align}
The dilaton and the metric equations of motion are given by:
\begin{align}
    & R = -2, \\
    \label{dilatoneqn}
    & \nabla_\mu \nabla_\nu \Phi - g_{\mu\nu} \nabla^2 \Phi + g_{\mu\nu} \Phi = -8\pi G_N T_{\mu\nu}.
\end{align}
 where \( T_{\mu \nu} \) is the stress-energy tensor of the matter sector. The dilaton equation of motion constrains the two-dimensional geometry to be locally AdS$_{2}$.
\subsection{Quantum Extremal Surfaces}

The entropy of Hawking radiation and the generalized entropy are given, respectively, by the following formulae: 
\begin{align}
S_R &= \min_I \text{Ext}_I
\left[
\frac{A(\partial I)}{4 G_N} + S_{\text{QFT}}(R \cup I)
\right], \label{island1} \\ \label{island2}
S_{\text{gen}} &= \frac{A(\partial I)}{4 G_N} + S_{\text{QFT}}(R \cup I).
\end{align}

Here, \( G_N \) denotes Newton's constant, \( R \) is the radiation region, \( I \) is the island, \( \partial I \) is the boundary of the island, and \( S_{\text{QFT}}(R \cup I) \) represents the entanglement entropy of the quantum fields in the bulk region of spacetime. The term \( R \cup I \) denotes the union of the island and the radiation region. Extremization of \( S_{\text{gen}} \) determines the location of the island. The surfaces that extremize the generalized entropy are called quantum extremal surfaces (QES).
\begin{figure}[H]
    \centering
    \begin{tikzpicture}[scale=1.5] 

\draw[thick] (-4,0) -- (-2,2); 
\draw[thick] (-2,-2) -- (-4,0); 
\draw[dashed, thick] (-2,-2) -- (0,0); 
\draw[dashed, thick] (0,0) -- (-2,2); 

\draw[thick] (4,0) -- (2,2); 
\draw[thick] (2,-2) -- (4,0); 
\draw[dashed, thick] (2,-2) -- (0,0); 
\draw[dashed, thick] (0,0) -- (2,2); 

\draw[thick, blue] (-2,-2.5) -- (-2,2.5); 
\draw[thick, blue] (2,-2.5) -- (2,2.5);   

\draw[ultra thick, black] (3, 0) -- (0.6, 0); 

\draw[black, thick] (2.9, 0) node[anchor=west] {$]$};  

\fill[black] (0.6, 0) circle (2pt); 

\draw[black] (0.6, -0.1) node[anchor=north] {\(-a\)}; 

\draw[black] (1.8, 0.2) node[anchor=south] {\(\Sigma_X\)}; 

\draw[->, thick, stealth-] (0.5
, 0.2) -- (0, 1.1) node[anchor=south] {\(X\)}; 

\draw[->, thick, stealth-] (3, 0.2) -- (3.5, 0.6) node[anchor=south] {\(Q\)};
\draw[black] (3, -0.1) node[anchor=north] {\(b\)};
\end{tikzpicture}
 
    \caption{Quantum Extremal Surface}
    \label{imagefirst}
\end{figure} 

When there is a non-trivial QES in the bulk, it corresponds to an island.
Figure \ref{imagefirst} shows a QES where the exact position of `a' is determined by extremizing the generalized entropy.

\subsection{Black Hole at Equilibrium }
We next introduce the setup of a black hole coupled to a bath. The framework of JT gravity black holes with leaky boundary conditions was originally developed in \cite{Engelsoy:2016xyb,Mertens:2019bvy}.

In our case, we have the AdS\(_2\) Poincar\'{e} patch with coordinates \(x^{\pm}\):
\begin{align}
ds^2 = -\frac{4 \, dx^{+} \, dx^{-}}{\left(x^{+} - x^{-}\right)^2}.
\end{align}where we have 
\[
u = \frac{x^{+} + x^{-}}{2} \in \mathbb{R}\quad\text{and} \quad s = \frac{x^{+} - x^{-}}{2} > {\epsilon}
\]

The AdS$_2$ boundary is parametrized by a coordinate transformation \( u = f(t) \) that relates the Poincar\'{e} coordinates $x^\pm$ to the Rindler coordinates $y^\pm$. The metric in the $y^\pm$ coordinates is given by:
\begin{align}
ds^2 =\frac{-4f'(y^+) f'(y^-)}{\left(f(y^+) - f(y^-)\right)^2} \, dy^+ \, dy^-
\end{align}
We attach a half-Minkowski thermal bath at the holographic time-like boundary at \( z = -\epsilon \). Note that this serves as our IR cutoff.
We use flat coordinates for the bath and we do so by analytically continuing the coordinates of the bulk.
\begin{equation}
ds^2_{bath} = \frac{-1}{\epsilon^2} \, dy^+ \, dy^-
\end{equation}

An eternal black hole with temperature \( \frac{1}{\beta} \) has the boundary parameterization relating the Poincar\'{e} and black hole patches given by:
\begin{eqnarray}
x^{\pm} = \frac{\beta}{\pi} \tanh \left( \frac{\pi y^{\pm}}{\beta} \right).
\label{5}
\end{eqnarray}
Then the black hole patch has a metric of the form:
\begin{eqnarray}
ds^2 = -\frac{\pi^2}{\beta^2} \frac{4}{\sinh^2\left( \frac{\pi}{\beta}(y^+ - y^-) \right)} \, dy^+ \, dy^-.
\end{eqnarray}

The ADM energy of spacetime is holographically related to the boundary parameterization:
\begin{eqnarray}
E(t) = -\frac{\Phi_{r}}{8 \pi G_N} \{f, t\}, \label{eq12}\\
\{f(t), t\} \equiv \frac{f'''(t)}{f'(t)} - \frac{3}{2} \left( \frac{f''(t)}{f'(t)} \right)^2. \label{eq13}
\end{eqnarray}
\(\{f(t), t\}\) is the Schwarzian  derivative of boundary parameterization.
In JT gravity, all nontrivial dynamics of the black hole is encoded in the boundary reparameterization mode $f(t)$. Since the bulk spacetime is locally AdS$_2$, the geometry itself carries no propagating degrees of freedom; instead, the physical dynamics arises from the way the boundary trajectory is embedded in AdS$_2$. The Schwarzian derivative $\{f,t\}$ determines the ADM energy via \eqref{eq12}, and hence the time dependence of $f(t)$ directly captures the evolution of the black hole mass and temperature.

At the boundary, the rate of change of energy is:
\begin{align}
\frac{\partial E}{\partial t}= T_{y^+y^+}-T_{y^-y^-}
\label{ADM}
\end{align}
Here \( T_{y^+ y^+} \) is the ingoing stress-energy tensor that originates from the bath attached to the black hole, and \( T_{y^- y^-} \) is the outgoing stress-energy tensor, i.e., the Hawking radiation from the black hole. We can always express the outgoing stress-energy tensor \( T_{y^- y^-} \) as a function of parameterization:
\begin{equation}
    T_{y^- y^-} =- \frac{c}{24 \pi} \{ f(y^-), y^- \} = k*E(y^-). \label{Adm simplified}
\end{equation}
 where $k=\frac{G_Nc}{3\Phi_r}$

Under a conformal transformation
\[ 
ds^2_g = \Omega^{-2} ds^2_\eta
\]
The stress-energy tensor components transform as follows: 
\begin{equation}
\left<T_{y^\pm y^\pm}\right>_g = \left<T_{y^\pm y^\pm}\right>_\eta - \frac{c}{12 \pi} \frac{\partial_\pm^2 \Omega}{\Omega},
\label{weyl}
\end{equation}

\begin{equation}
\left( \frac{\partial x^\pm}{\partial y^\pm} \right)^2 \left<T_{x^\pm x^\pm}\right>_\eta = \left<T_{y^\pm y^\pm}\right>_\eta + \frac{c}{24} \{x^\pm, y^\pm\}.
\label{coord weyl}
\end{equation}
Equation \eqref{weyl} describes a Weyl transformation of stress-energy tensors, here $\left<T_{y^\pm y^\pm}\right>_g $ corresponds to the normal ordered stress tensor with respect to the frame with metric $ds^2 = -\Omega^2(y^+,y^-)\, dy^+dy^-$, and $\left<T_{y^\pm y^\pm}\right>_\eta $ corresponds to the normal ordered stress tensor with respect to the frame with metric $ds^2 = -dy^+dy^-$.
Equation \eqref{coord weyl} represents a complete conformal transformation, where, $\left<T_{x^\pm x^\pm}\right>_\eta $ corresponds to the normal ordered stress tensor with respect to the flat metric $ds^2 = - dx^+dx^-$, and $\left<T_{y^\pm y^\pm}\right>_\eta $ corresponds to the normal ordered stress tensor with respect to the flat metric $ds^2 = -dy^+dy^-$ .

A stationary black hole corresponds to an exponential form of $f(t)$, for which the Schwarzian is constant and the energy is fixed. Both the incoming and the outgoing modes of the stress tensor have temperature $\frac{1}{\beta}$
\begin{equation}
    \left<T_{y^\pm y^\pm}\right> = \frac{\pi c}{12\beta^2}, \label{seo}
\end{equation}
Note that the expectation values from this point onward are normal ordered stress tensors with respect to a flat metric in the respective coordinate. We remove the subscript from the further discussion.\\
In equilibrium, we have 
\begin{equation}
    \frac{\partial E}{\partial t} = 0
\end{equation}
\\
We can transform our stress tensor value using \eqref{coord weyl}, and in doing so we get 
    $\left<T_{x^\pm x^\pm}\right> = 0$, which is the expected result.
    
\subsection{Entropy in terms of boundary parameterization}
First, we introduce a new frame $w^\pm$ where the CFT is in a vacuum state with a flat background geometry: 
\begin{equation}\label{se1}
    \left\langle T_{w^{ \pm} w^{ \pm}}\right\rangle=0 
\end{equation}

In this frame, the matter contribution to entanglement entropy between two points $a$ and $b$ is given by the Cardy-Calabrese formula \cite{Calabrese:2004eu}:
\begin{equation}\label{cardy}
    S=\frac{c}{6}\log\left|\frac{(w^+_a-w^+_b)(w^-_a-w^-_b)}{\epsilon^{w^+_a}\epsilon^{w^-_a}\epsilon^{w^+_b}\epsilon^{w^-_b}}\right|.
\end{equation} 
Here, ${\epsilon^{w^+_a},\epsilon^{w^-_a},\epsilon^{w^+_b},\epsilon^{w^-_b}}$ are the UV cutoffs for the entropy at points $a$ and $b$. The cutoffs are taken with respect to the proper distance in the physical metric.\\

Next, we want to relate the $w\pm$frame to the original coordinate system $y\pm$. The stress energy tensor in the two frames is related by:
\begin{equation}\label{serel}
    \left(\frac{d w^{ \pm}}{d y^{ \pm}}\right)^2\left\langle T_{w^{ \pm} w^\pm}\right\rangle=\left\langle T_{y^\pm y^{ \pm}}\right\rangle+\frac{c}{24 \pi}\left\{w^\pm, y^{ \pm}\right\}
\end{equation}

First, consider the coordinate $w^+$.

We have from equations \eqref{se1} and \eqref{serel}:
\begin{align}
&\left\{w^{+}, y^{+}\right\}=-\frac{24 \pi}{c}\left\langle T_{y^{+} y^{+}}\right\rangle, 
\end{align}
Combining this with \eqref{seo}, we get:
\begin{align} 
\label{bef}
& \left\{w^{+}, y^{+}\right\}=-\frac{2 \pi^2}{\beta^2}, \quad y^{+}<0,
\end{align}
After $y^+ \geq 0$ the incoming bath stress tensor changes so we have:
\begin{align} \label{aft}
& \left\{w^{+}, y^{+}\right\}=-\frac{2\pi^2}{{\beta_{bath}}^2}, \quad y^{+} \geqslant 0
\end{align}

During $y^+ <0$, $w^+=e^{\frac{2\pi}{\beta}y^+}$ solves \eqref{bef} and provides a coordinate system for both the black hole and the bath. After $y^+=0$, the coordinate relation between $y^+$ and $w^+$ is to be obtained by solving \eqref{aft}. 

As \eqref{aft} is a third-order differential equation, we have to specify three initial conditions at $y^+=0$. We impose continuity in the $w^+$ coordinates up to the second-order derivative.

This choice comes from the fact that a discontinuity in $w,w^\prime$ or $w^{\prime\prime}$ would correspond to unphysical discontinuities in the geometry, while the change (may be discontinuous) in the third derivative is precisely what encodes the change (may be instantaneous) in energy flux from the bath.

This matches the physical expectation that a sudden temperature change should not introduce discontinuities in the geometry. Rather, it should produce a finite change in the stress tensor flux.

This gives us:
\begin{equation}
\begin{aligned}
w^+|_{y^+=0}&=1\\
\frac{dw^+}{dy^+}\Big|_{y^+=0}&=\frac{2\pi}{\beta}\\
\frac{d^2w^+}{dy^{+2}}\Big|_{y^+=0}&=\frac{4\pi^2}{\beta^2}
\end{aligned}
\end{equation}

For the case of discontinuous change in the bath temperature, the discontinuity is reflected as the discontinuity of the Schwarzian in the third-order derivative. This discontinuity is an artifact of our rather artificial assumption of a discontinuous change in the bath temperature.

A similar argument follows for $w^-$ coordinates, where we have
\begin{align}
& \left\langle T_{y^-y^-}\right\rangle=\frac{\pi c}{12 \beta^2}, \quad y^{-}<0, &\left\langle T_{y^-y^-}\right\rangle=g(t) \geqslant 0
\end{align}
Here, $g(t)$ is obtained by solving \eqref{ADM} for a specific bath cooling profile. It follows that for $y^{-}<0$, the relation between $y^-$ and $w^-$  has the same form as \eqref{bef}:
\begin{align}
\left\{w^{-}, y^{-}\right\}= & -\frac{2 \pi^2}{\beta^2}, \quad y^{-}<0
\end{align}
This is solved by $w^-=-e^{\frac{-2\pi}{\beta}}$.\\
For $y^{-}\geq0$ the form of the relation is different from \eqref{aft}, however: 
\begin{align}
\left\{w^{-}, y^{-}\right\}=&  g(y^-) \geqslant 0 \label{w-}
\end{align}\\
Here too, we choose initial conditions that ensure continuity up to the second-order derivative:\\
\begin{equation}
\begin{aligned}
w^-|_{y^+=0}&=-1\\
\frac{dw^-}{dy^-}\Big|_{y^+=0}&=\frac{2\pi}{\beta}\\
\frac{d^2w^-}{dy^{-2}}\Big|_{y^+=0}&=-\frac{4\pi^2}{\beta^2}
\end{aligned}
\label{bc1}
\end{equation}

Unlike in the previous case, here the Schwarzian is never discontinuous. This reflects the fact that although the bath temperature may change suddenly, the black hole temperature changes in a continuous manner.

The cut-off distances in \eqref{cardy} are taken in terms of the physical metric, which is $\epsilon^{y^\pm}$. In the w-frame, these are given by:
\begin{align}
    \epsilon^{w^+}\epsilon^{w^-} &= -dw^+dw^-\\
    -dw^+dw^-&=-dy^+dy^- \left(\frac{dy^+}{dw^+}\right)\left(\frac{dy^-}{dw^-}\right)\\
    -dw^+dw^-&=-\Omega^2dy^+dy^- \left(\frac{dy^+}{dw^+}\right)\left(\frac{dy^-}{dw^-}\right)\frac{1}{\Omega^2}\\
    -dw^+dw^-&= \epsilon^{y^+}\epsilon^{y^-}\left(\frac{dy^+}{dw^+}\right)\left(\frac{dy^-}{dw^-}\right)\frac{1}{\Omega^2}\\
    \epsilon^{w^+}\epsilon^{w^-}&= \epsilon^{y^+}\epsilon^{y^-}\left(\frac{dy^+}{dw^+}\right)\left(\frac{dy^-}{dw^-}\right)\frac{1}{\Omega^2}
\end{align}
The blackening factor $\Omega^2$ for the metric in the y-frame is given by: 
\begin{equation}
\Omega^2=\frac{-4f'(y^+) f'(y^-)}{\left(f(y^+) - f(y^-)\right)^2}
\end{equation} 
where $f$ is the boundary parameterization function. Note that for the point $a$ in the bulk, $\Omega^2$ varies, but for the point $b$ in the bath, it is a constant. 

Next, we fix the dilaton. In the semi-classical limit, the dilaton is obtained by solving the dilaton equation \eqref{dilatoneqn}. We solve it in the $x^\pm$ frame and then relate the result to the $y^\pm$ frame using $x^\pm=f(y^\pm)$. To do this, we first determine $\left\langle T_{x^\pm x^\pm}\right\rangle$. From \eqref{coord weyl}, we observe that $\left\langle T_{x^+ x^+}\right\rangle$ and $\left\langle T_{x^- x^-}\right\rangle$ both vanish. The non-zero component $\left\langle T_{x^+ x^+}\right\rangle$ is given by:
$$
T_{x^{+} x^{+}}=\frac{1}{f^{\prime}\left(y^{+}\right)^2}\left(\frac{\pi c}{12 \beta^2}+\frac{c}{24 \pi}\left\{f\left(y^{+}\right), y^{+}\right\}\right) .
$$
From \eqref{ADM}, this equals:
$$
T_{x^{+} x^{+}}=\frac{1}{f^{\prime}\left(y^{+}\right)^2} \partial_{y^{+}} E\left(y^{+}\right)=-\frac{\Phi_r}{8 \pi G_N} \frac{1}{f^{\prime}\left(y^{+}\right)^2} \partial_{y^{+}}\left\{f\left(y^{+}\right), y^{+}\right\}
$$
Using $\left\{f^{-1}\left(x^{+}\right), x^{+}\right\}=-\left(f^{\prime}\left(y^{+}\right)\right)^{-2}\left\{f\left(y^{+}\right), y^{+}\right\}$, we get:
$$
T_{x^{+} x^{+}}=-\frac{\Phi_r}{8 \pi G_N} \partial_{x^{+}}^3 f^{\prime}\left(y^{+}\right)
$$\\
Next, we follow the argument as in \cite{Hollowood:2020cou} to obtain the expression for the dilaton:
$$
\Phi=\Phi_0+2 \Phi_r\left(\frac{f^{\prime \prime}\left(y^{+}\right)}{2 f^{\prime}\left(y^{+}\right)}+\frac{f^{\prime}\left(y^{+}\right)}{f(y^{-})-f\left(y^{+}\right)}\right)
$$\\

Next, we calculate the entropy of the Hawking saddle and Island saddle.\\

 \section{Dynamical Setup and Numerical Framework}
In this section, we will set up the non-equilibrium problem of a black hole interacting with a bath of changing temperature. We will introduce our numerical methods and use them to obtain the Page curve for the case of an instantaneous change in temperature. We will see that our numerical results match with the analytical results obtained by \cite{Hollowood:2020cou} to a very high degree of accuracy. This serves as a check of our numerical methods.

\subsection{Breaking equilibrium: time-dependent bath temperature}
When the incoming stress tensor from the bath is modified, the energy balance equation drives the Schwarzian away from its constant value. The resulting time dependence of $f(t)$ describes the relaxation of the black hole from one thermal state to another.

Thus, the non-equilibrium evaporation problem reduces to solving for the dynamics of this single boundary degree of freedom. Once $f(t)$ is known, the dilaton profile, the Hawking flux, and the location of the quantum extremal surfaces are all determined. The evolution of the Page curve is therefore controlled entirely by the dynamics of the Schwarzian mode.

The equilibrium in \eqref{ADM} is broken when the incoming stress tensor from the bath is modified in any manner, leading to the following. 
\begin{align}
\frac{\partial E}{\partial t}= T_{y^+y^+}^{\text{modified}}-T_{y^-y^-}
        \label{ADM2}
\end{align}
Using \eqref{Adm simplified} in \eqref{ADM2} we get the following:
\begin{align}
\frac{\partial E}{\partial t}+k.E= T_{y^+y^+}^{\text{modified}}
        \label{ADM3}
\end{align}
Equation \eqref{ADM3} shows that the entire non-equilibrium evolution is driven by the time-dependent incoming flux from the bath. The function $T^{\text{modified}}_{y^+y^+}(t)$ is the only external input in the problem. Once this flux is specified, the black hole energy $E(t)$ is determined by solving the first-order differential equation \eqref{ADM3}, with $k$ setting the intrinsic relaxation timescale.

Through the relation \eqref{eq12}:
\[
E(t) = -\frac{\Phi_r}{8\pi G_N}\{f,t\},
\]
the time dependence of the energy directly fixes the Schwarzian derivative of the boundary reparameterization. The problem therefore reduces to solving a third-order differential equation for $f(t)$ with appropriate initial conditions. In general, for arbitrary time-dependent flux, this equation does not admit an analytic solution and must be solved numerically.

Thus, the dynamical content of the evaporation process is entirely encoded in the specified incoming stress tensor. Different physical situations correspond simply to different choices of $T^{\text{modified}}_{y^+y^+}(t)$, while the governing equations remain the same.

This implies that all situations where the stress tensor is modified in a time dependent way are mathematically equivalent. We study a cooling bath, but the method applies equally to, for instance, a situation where negative energy radiation is emitted from the black hole to the bath. Another example would be the partially absorbing boundary condition. 
Since evolution depends on a single parameter $T_{y^+y^+}^{\text{modified}}$, all of these different cases can be captured by the same differential equations \eqref{ADM3} and \eqref{eq12}.


\subsection{Instantaneous change in bath temperature: Boundary parameterization}
We next consider the case of an instantaneous change in the bath temperature. We start with the bath and the black hole in equilibrium at inverse temperature $\beta$ and then make a sudden change at boundary time $t=0$. Before the change, at $t<0$ or alternatively \(y^+<0\), we have:
\begin{equation}
    T_{y^+ y^+} = \frac{\pi c}{12 \beta^2}
\end{equation}
The temperature of the bath instantaneously changes to $\tilde{\beta}$ at the boundary time $t=0$.
This change propagates causally to the bulk. 
After \(y^+>0\), we get :
\begin{equation}\label{seo2}
    T_{y^+ y^+} = \frac{\pi c}{12 \tilde{\beta}^2}
\end{equation}

Note that this setup is mathematically completely equivalent to the one studied in \cite{Hollowood:2020cou}, where a positive energy shock wave was sent to the black hole from the bath and the temperature of the black hole shifts and settles down to a new temperature. As a consistency check on our numerical methods, we will show that our results for boundary parameterization match the analytical results of \cite{Hollowood:2020cou}. 
    
Energy conservation at the holographic boundary \eqref{ADM3} gives:
\begin{equation}\label{Admenergy}
\frac{\partial E}{\partial t}+k.E = \frac{\pi c}{12 \tilde{\beta}^2}
\end{equation}\\
Solving the first-order differential equation, we get:
\begin{eqnarray}
    E=e^{-kt}\int \left[\frac{\pi c}{12\tilde{\beta^2}}e^{kt}dt \right] \label{eq43}
\end{eqnarray}\\
\\ 
We solve this differential equation with the boundary condition that at $t=0$ the ADM energy is $E=\frac{\pi\Phi_r}{4\beta^2G_N}$, which was the original energy of the black hole before any evaporation took place. Upon substituting `E' in the above equation\eqref{eq43} with $\{f,t\}$ using \eqref{eq12}, the third-order differential equation for the boundary parameterization is: 
\begin{equation}\label{similar}
 \{f,t\}=-2\pi^2[\tilde{\beta}^{-2}-(\tilde{\beta}^{-2}-\beta^{-2})e^{-kt}].   
\end{equation}
At early times, the black hole was in equilibrium with the bath at temperature $T_\beta$, so:
\begin{align}
   & f(y^\pm) = \frac{\beta}{\pi} \tanh\left(\frac{\pi}{\beta} y^\pm\right) \\
   & f'(y^\pm) = \text{sech}^2\left(\frac{\pi}{\beta} y^\pm\right) \\
   & f''(y^\pm) = -\frac{2\pi}{\beta} \sinh\left(\frac{\pi}{\beta} y^\pm\right) \, \text{sech}^3\left(\frac{\pi}{\beta} y^\pm\right)
\end{align}\\
At very late times, it once again settled to equilibrium at temperature $T_{\tilde{\beta}}$, so:
\begin{align}
    & f(y^\pm) \approx \frac{\tilde{\beta}}{\pi} \tanh\left( \frac{\pi}{\tilde{\beta}} y^\pm \right) \\
    & f'(y^\pm) = \text{sech}^2\left( \frac{\pi}{\tilde{\beta}} y^\pm \right) \\
    & f''(y^\pm) = -\frac{2\pi}{\tilde{\beta}} \sinh\left( \frac{\pi}{\tilde{\beta}} y^\pm \right) \text{sech}^3\left( \frac{\pi}{\tilde{\beta}} y^\pm \right)
\end{align}\hfill\\
 The initial conditions are imposed to ensure the continuity of $f(t)$ up to the second derivative.
 
\begin{equation}
\begin{aligned}
f|_{t_0=0}&=0\\
\frac{df}{dt}\Big|_{t_0=0}&=1\\
\frac{d^2f}{dt}\Big|_{t_0=0}&=0
\end{aligned}
\label{bc2}
\end{equation}

Note that in general, $\{f,t\} = g(t)$, and for the single instant drop in bath temperature $g(t)= -2 \pi^2\left[\tilde{\beta}^{-2}-\left(\tilde{\beta}^{-2}-\beta^{-2}\right) e^{-k t}\right]$. In general $g(t)$ would have to be solved from \eqref{ADM}. The differential equation to solve for `f' is the same as $w^-$\eqref{w-}, but with a slightly different initial condition as seen in \eqref{bc1} and \eqref{bc2}. This makes them related by a M\"obius transformation \cite{Hollowood:2020cou}. 

To ensure that the back-reaction on the geometry remains under control and that the semiclassical approximation is valid, we must have $k<<1$. 

The exact solution of this differential equation was obtained in \cite{Almheiri:2019qdq,Hollowood:2020cou}
where $f(t)$ is expressible in terms of modified Bessel functions. The exact analytical solution for boundary parameterization is as follows: 
\[
f(t) = \frac{\tilde\beta}{\pi}\frac{  K_{\nu}(\nu z_0) \left( \frac{\hat{f}(t)}{\hat{f}(t_0)} - 1 \right) + z_0 \tanh \frac{\pi t_0}{\tilde{\beta}} \left( \hat{f}(t) I'_{\nu}(\nu z_0) / \alpha - K'_{\nu}(\nu z_0) \right)}
{K_{\nu}(\nu z_0) \left( \frac{\hat{f}(t)}{\hat{f}(t_0)} - 1 \right) \tanh \frac{\pi t_0}{\tilde{\beta}} + z_0 \left( \hat{f}(t) I'_{\nu}(\nu z_0)/\alpha - K'_{\nu}(\nu z_0)\right)}
\]

where
\[
\hat{f} = \alpha \frac{K_\nu(\nu z)}{I_\nu(\nu z)}, \qquad
\nu = \frac{2\pi}{\tilde{\beta} k}, \qquad
z = \sqrt{\frac{E_{\text{shock}}}{E_\beta}} \, e^{-k(t-t_0)/2}.
\]
The constant of normalization is fixed by using the initial condition on $\hat{f(t_0)}=e^{2\pi t_0/\beta}$. Also for our case $t_0=0$. Here, the AdS radius is set to 1. 

We next proceed to solve this equation numerically.  The time step for evolution is taken to be 0.01. We use the RK4 method to numerically solve the differential equation to obtain the boundary parameterization $f(t)$ and the `$w\pm$' coordinates.

\begin{figure}[H]
\centering
\begin{subfigure}{0.45\textwidth}
    \centering
    \includegraphics[width= \textwidth]{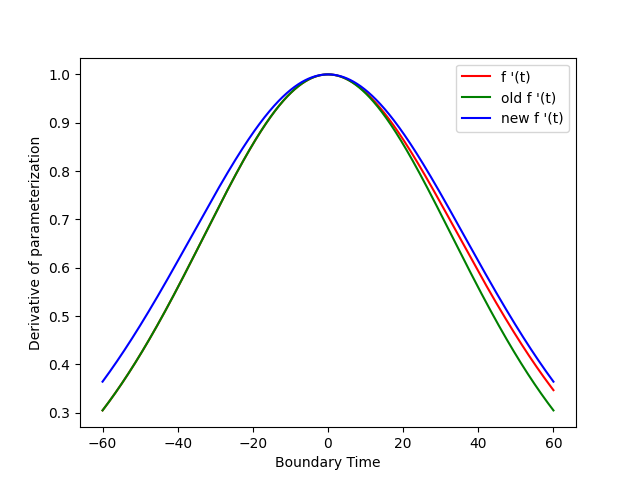}
    \caption{Plot of first derivative of $f(t)$ with $t$}
    \label{df}
\end{subfigure}
\hfill
\begin{subfigure}{0.45\textwidth}
    \centering\textbf{}
    \includegraphics[width= \textwidth]{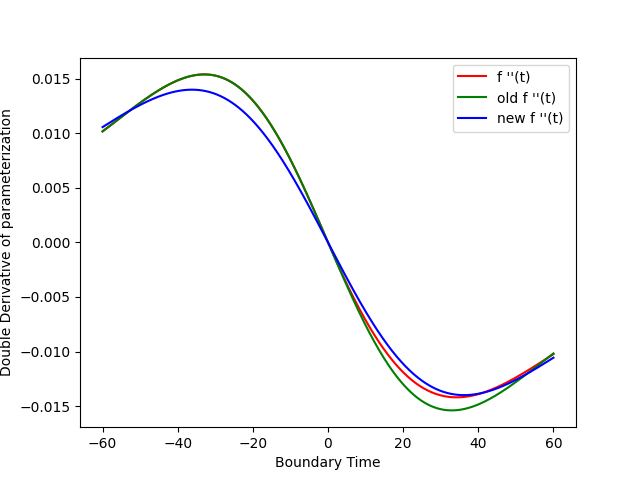}
    \caption{Plot of second derivative of $f(t)$ with $t$}
    \label{ddf}
\end{subfigure}
\caption{Interpolation of derivatives}
\small{Red curves show numerical RK4 solution, green shows analytical early-time (equilibrium at $\beta$), blue shows analytical late-time (equilibrium at $\tilde\beta$)}
\end{figure}

First, we plot the first and second derivatives of $f(t)$. This is shown in Fig \ref{df} and Fig \ref{ddf}. As a consistency check, we match them with the analytically computed derivatives of $f(t)$ at early and late times.

In Figures \ref{df} and \ref{ddf}, the numerically computed derivative is shown in red, while the analytically computed early and late time derivatives are shown in green and blue, respectively. The figure demonstrates how the derivative and double derivative of the boundary parameterization interpolate between the old and new boundary parameterizations.
\begin{figure}[H]
\begin{subfigure}{0.45\textwidth}
    \centering
    \includegraphics[width= \textwidth]{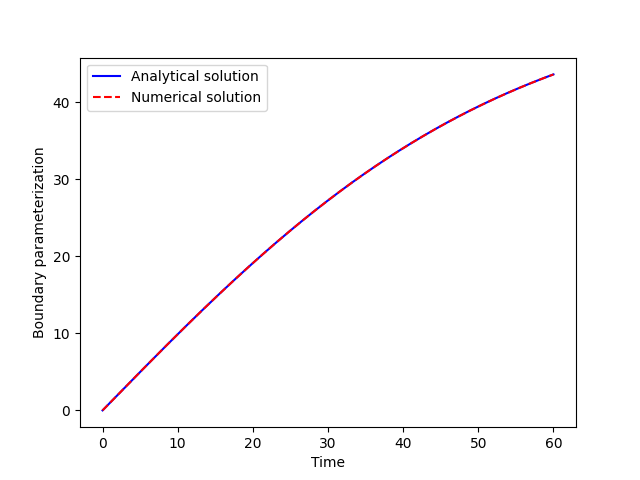}
    \caption{Plot of numerical and analytical f(t)}
    \label{analytical}
\end{subfigure}
\hfill
\begin{subfigure}{0.45\textwidth}
    \centering\textbf{}
    \includegraphics[width= \textwidth]{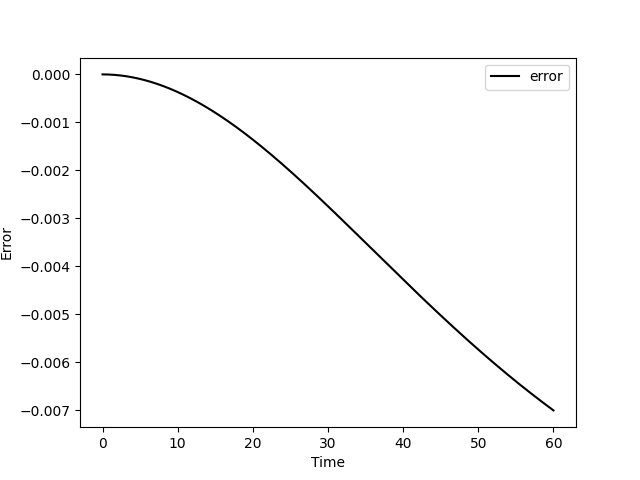}
    \caption{plot of cumulative error between analytical and numerical f(t)}
    \label{error}
\end{subfigure}
\caption{Numerical check with analytical result}
\end{figure}

Figure \ref{analytical} is a comparative plot between our numerical solution and the analytical solution of \cite{Hollowood:2020cou}. The numerical and analytical solutions agree within marginal accumulation error as shown in Fig.\ref{error}. The accumulated error remains below $10^{-4}$, validating our numerical RK4 method.

RK4 typically has an error rate of the order $O(h^4)$, where h (=0.01 in our case) is the time step. Our results are insensitive to this level of numerical error; in fact, the standard Euler method for numerical integration also gives the same result. In order to satisfy the semiclassical regime, our k is set to 0.1 throughout the discussion. Other parameters include $\beta=50\pi$ and $\tilde\beta=55\pi$, $c=100$ and $G_N=0.06$.

\subsection{Instantaneous Change in Temperature: Entropy saddles}
Having computed the evolution of the boundary parameterization with time, we proceed to use it to compute the generalized entropy in the Hawking saddle and the Island saddle. 

\subsubsection{Hawking saddle}
 We begin with the calculation of the Hawking saddle.
\begin{figure}[H]
    \centering
    \begin{tikzpicture}[scale=1.2, xscale=1] 

\draw[thick] (-4,0) -- (-2,2); 
\draw[thick] (-2,-2) -- (-4,0); 
\draw[dashed, thick] (-2,-2) -- (0,0); 
\draw[dashed, thick] (0,0) -- (-2,2); 

\draw[thick] (4,0) -- (2,2); 
\draw[thick] (2,-2) -- (4,0); 
\draw[dashed, thick] (2,-2) -- (0,0); 
\draw[dashed, thick] (0,0) -- (2,2); 

\draw[thick, green] (-2.2,0) -- (2.2,0); 
\draw[green, thick] (-2,0) node[anchor=east] {\([\,\)} (-2.3,0)node[below] {\(b_L\)}; 
\draw[green, thick] (2,0) node[anchor=west] {\(\,]\)} (2.4,0)node[below] {\(b_R\)}; 

\draw[thick, blue] (-2,-2.5) -- (-2,2.5); 
\draw[thick, blue] (2,-2.5) -- (2,2.5);   

\draw[thick, blue] (-2.2,1) -- (2.2,1); 

\draw[blue, ultra thick] (-2, 1) node[anchor=east] {\([\,\)} (-2.3,1)node[below] {\(b_L\)};  
\draw[blue, ultra thick] (2, 1) node[anchor=west] {\(\,]\)}(2.4,1)node[below] {\(b_R\)};  

\end{tikzpicture}
    \caption{Hawking saddle represented geometrically. Entropy is related to the proper distance between the two points.}
    \label{hawking saddle}
\end{figure}
To compute the entropy, we first determine the $w^\pm$ coordinates for both endpoints $b_L$ and $b_R$ (as shown in Figure \ref{hawking saddle}). We start by specifying the points in the y frame and then write them in the w-frame coordinates. In the y-frame, we fix the spatial distance of the points from the cutoff surface. We set $z=2\epsilon$, so at any boundary time t the $y^\pm$ coordinates of the points in the right bath are:
\begin{align}
    &y^+=(t+z)/2, \quad &y^-=(t-z)/2
\end{align}
We proceed similarly for the left side bath and get these points in the corresponding $w^\pm$ coordinates. Note that the $w^\pm$ coordinates that we use to completely span the two-sided black hole, along with the two baths attached on either side. Then the Hawking saddle is the entanglement entropy between the two bath points using the Cardy-Calabrese formula \eqref{cardy}.\\
\begin{figure}[H]
    \centering
    \includegraphics[scale=0.8]{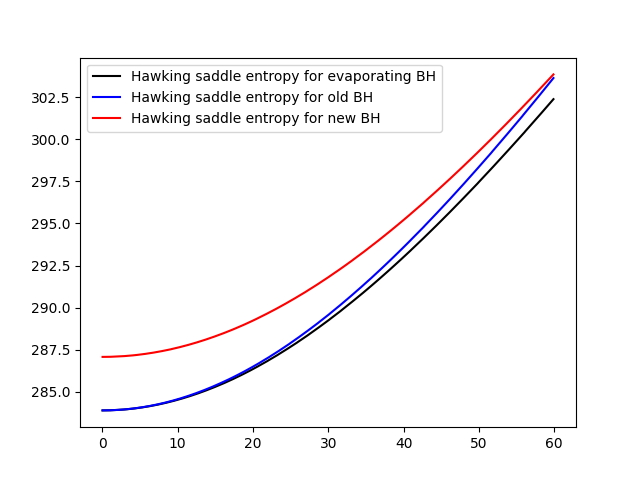}
    \caption{Entropy of Hawking saddle vs Boundary time.}
    \label{hawking saddle entropy}
\end{figure}
In Figure \ref{hawking saddle entropy}, the entropy curves corresponding to the old and new black holes are the Hawking saddle entropy curves for eternal black holes corresponding to the initial and final temperatures, respectively. The entropy curve for our evaporating black hole starts as the old one but later becomes parallel to the new saddle. As expected, the Hawking saddle entropy curves become linear at late times, with slope proportional to the temperature of the black hole. The entropy plot for the evaporating black hole thus shows an interpolation between the initial and final states.
\subsubsection{Island saddle}
\begin{figure}[H]
    \centering
    \begin{tikzpicture}[scale=1.3] 

\draw[very thick] (0,0) -- (3,3) -- (0,6) -- (-3,3) -- cycle;

\draw[very thick, blue] (-1.5,0) -- (-1.5,6); 
\draw[very thick, blue] (1.5,0) -- (1.5,6);   

\coordinate (A) at (-1.5,1.5); 
\coordinate (B) at (-1.5,4.5); 
\coordinate (C) at (1.5,1.5);  
\coordinate (D) at (1.5,4.5);  

\draw[dashed, very thick] (A) -- (D); 
\draw[dashed, very thick] (B) -- (C); 

\fill[red] (2,3) circle (2pt) node[above] {\(w(y_1)\)} node[below] {\(b\)}; 

\fill[red] (0.4,3) circle (2pt) node[above] {\(w(y_2)\)} node[below] {\(a\)};;  
\fill[red] (-2,3) circle (2pt)  node[below] {\(b\)}; 

\fill[red] (-0.4,3) circle (2pt)  node[below] {\(a\)};;  
\draw[red, <->, very thick] (2,3) -- (0.4,3);
\draw[red, <->, very thick] (-2,3) -- (-0.4,3);
\end{tikzpicture}
    \caption{Geometric representation of island saddle. Entropy is related to the proper distance along the red lines and the value of dilaton at `a'. The points `a' are QES with respect to the points `b'.}
    \label{island saddle}
\end{figure}
In the calculation of island saddle, we iterate over all possible points `a' (`$y_2$' in Figure \ref{island saddle}) in a fixed time slice to find an extremum. Note that there is a symmetry between the left and right sides of the diagram.
\begin{figure}[H]
    \centering
    \includegraphics[width=0.75\textwidth]{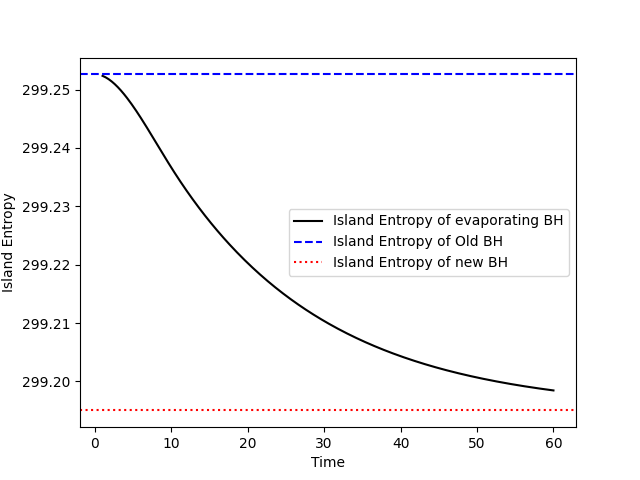}
    \caption{Entropy of Island saddle vs Boundary time.}
    \label{island entropy}
\end{figure}

Figure \ref{island entropy} shows the decrease in the island entropy over time from the black hole entropy corresponding to the black hole at the initial temperature to that corresponding to a black hole at the new temperature. The asymptotic behavior of the black hole coming to equilibrium in its final state is clearly reflected in the graph. The Page curve comes down for an evaporating black hole, whereas it stays at a constant value for an eternal black hole. Thus, we have directly obtained the Page curve after Page time for an evaporating black hole. In addition, this numerical approach can be used to profile evaporation for any type of cooling that the bath might provide. 

\section{Black Hole Evaporation: General Cooling Profiles}
In this section, we analyze different examples of cooling of the bath, which are continuous in nature and different from the earlier instantaneous drop in temperature.

 The starting point for the general case is the generalization of \eqref{ADM3}:
\begin{equation}\label{Admenergynew}
\frac{\partial E}{\partial t} + k.E =T^{modified}_{y^+y^+} = \frac{\pi c}{12 \tilde{\beta_{t}}^2} 
\end{equation}\\
where $\tilde{\beta_{t}}^2$ is the temperature of the bath, which is now a general time dependent function. Once again, $T^{modified}_{y^+y^+}$ is our only input. Now we solve the differential equation \eqref{Admenergynew}, we obtain the energy $E(t)$ of the black hole. As before, we solve\eqref{eq12} and \eqref{eq13} to get the boundary parameterization ($f(t)$) and use it to compute the entropies for the Hawking and the island saddles.

We will focus on three examples: a case where the bath temperature follows Newton's law of cooling, one where the incoming stress tensor from the bath has a drop in value linear in time, and finally, a quasi-static case.

\subsection{Bath Energy follows Newton's law of cooling}

Newton's law of cooling corresponds to the natural relaxation of a finite bath toward equilibrium with an external reservoir. It is widely used as an effective description in statistical mechanics.

In this case, we take $\frac{\pi c}{12\tilde{\beta_{t}}^2} = \frac{\pi c}{12}\left(\frac{1}{\tilde\beta^2}+\left(\frac{1}{\beta^2}-\frac{1}{\tilde\beta^2}\right)e^{-st}\right)$. We choose `$s$' as the cooling constant. We first solve for `$E$' from the first-order differential equation and then solve for the boundary parameterization from the Schwarzian. The same numerical procedure as the previous section is applied, now replacing $\tilde{\beta}$ with $\tilde{\beta_t}$.

We analyze how both the black hole entropy and the island entropy evolve for different relationships between the cooling constant $s$ and the parameter $k$. The three cases of interest are $k > s, k < s,$ and $k = s$ which represent different rates of cooling. In each case, we also monitor the evolution of the black hole energy. For concreteness, we present numerical results for two representative choices: $s=k/2$ and $s=2k$.

\begin{figure}[H]
    \centering
    \begin{subfigure}{0.4\textwidth}
    \centering
    \includegraphics[width=1\linewidth, height=0.2\textheight]{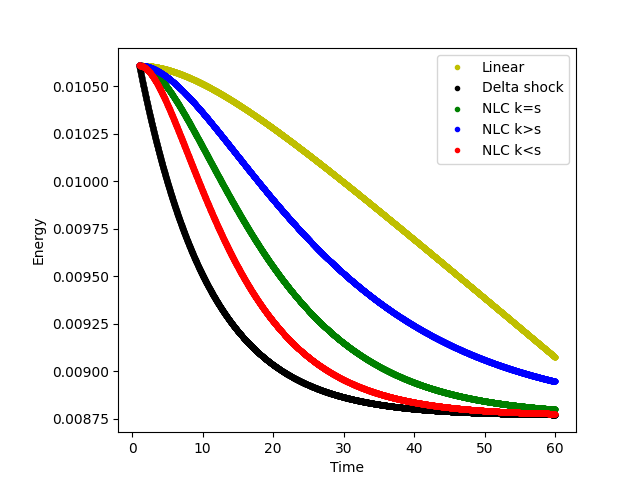}
    \caption{Black hole energy vs Time.}
    \label{figure8a}
\end{subfigure}
\begin{subfigure}{0.4\textwidth}
    \centering
    \includegraphics[width=1\linewidth, height=0.2\textheight]{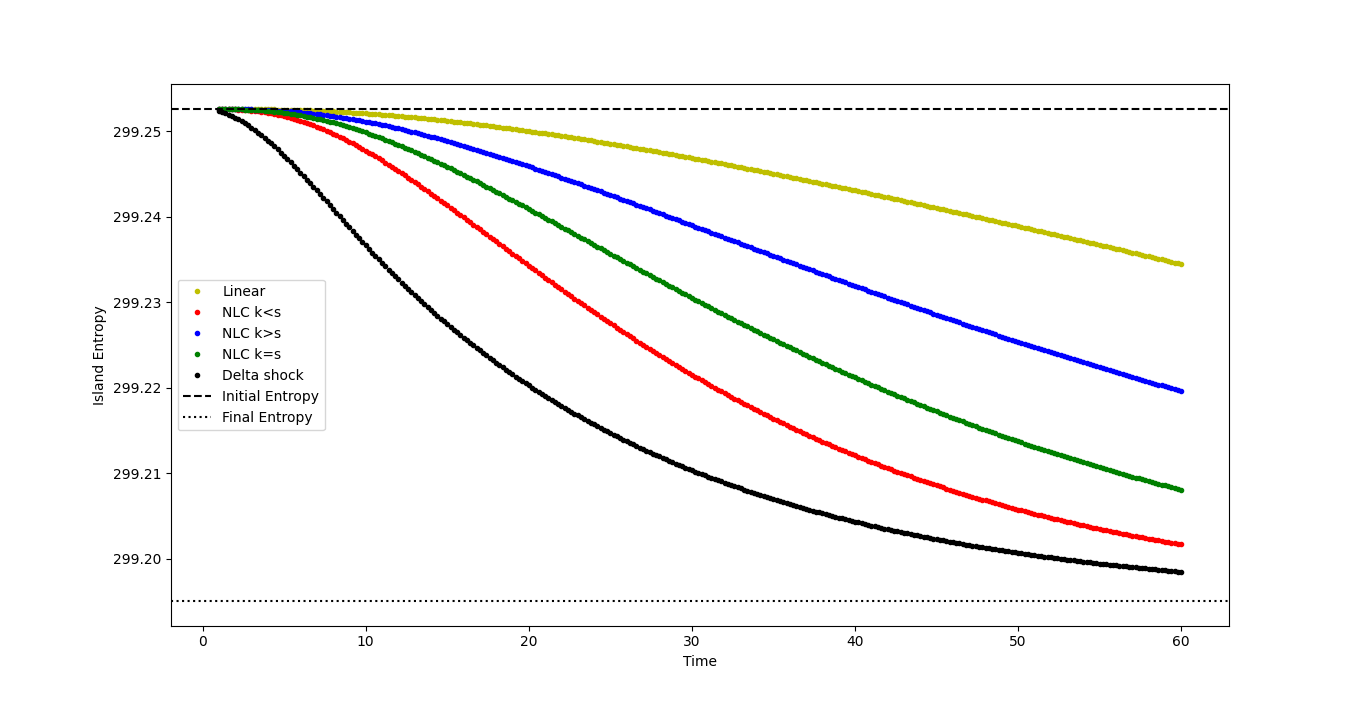}
    \caption{Island entropy vs Time.}
    \label{figure8b}
\end{subfigure}
\caption{\small The different colors represent different bath cooling profiles. Yellow: bath temperature decreasing linearly, Black: a single instant drop in bath temperature, Red/Green/Blue: bath temperature follows Newton's law of cooling (NLC) for $(k < s), (k = s), \text{ and } (k > s)$ cases, respectively.}
\end{figure}
In the figures \ref{figure8a} and \ref{figure8b} we observe the change in energy of the black hole and its island saddle entropy, respectively. The Island entropy curve is seen to track the energy curve. This is consistent with the decrease in black hole energy and consequently the area of the black hole. The island entropy at QES corresponds to the area of the black hole horizon (dilaton is our case) which decreases as the black hole evaporates, thus a dynamically computed entropy always tracks the energy.

\begin{figure}[H]
\centering
\begin{subfigure}{0.4\textwidth}
    \centering
    \includegraphics[width=1\linewidth, height=0.2\textheight]{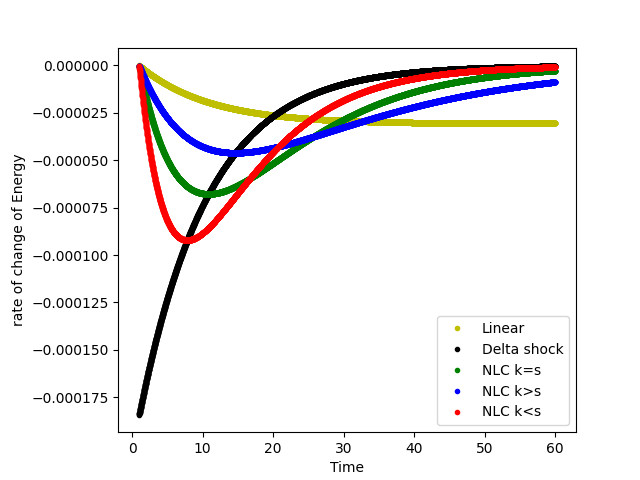}
    \caption{Rate of change of energy.}
    \label{figure9a}
\end{subfigure}
\begin{subfigure}{0.4\textwidth}
    \centering
    \includegraphics[width=1\linewidth, height=0.2\textheight]{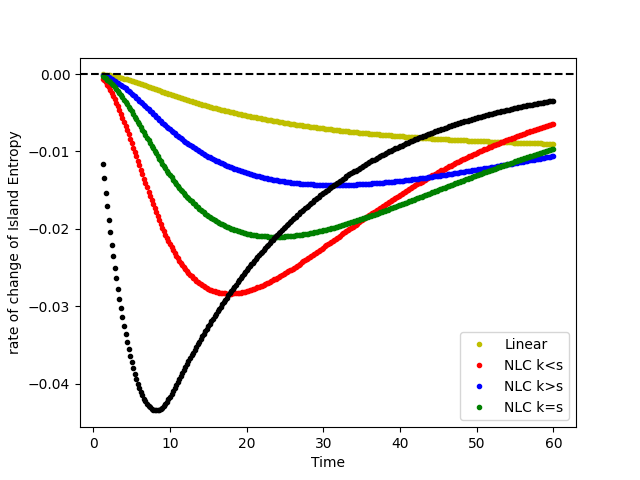}
    \caption{Rate of change of island entropy.}
    \label{figure9b}
    \end{subfigure}
    \caption{{\small The different colors represent different bath cooling profiles. Yellow: bath temperature decreasing linearly, Black: a single instant drop in bath temperature, Red/Green/Blue: bath temperature follows Newton's law of cooling (NLC) for $(k < s), (k = s), \text{ and } (k > s)$ cases, respectively.}}
\end{figure}

\begin{figure}[H]
    \centering
    \includegraphics[width=\linewidth]{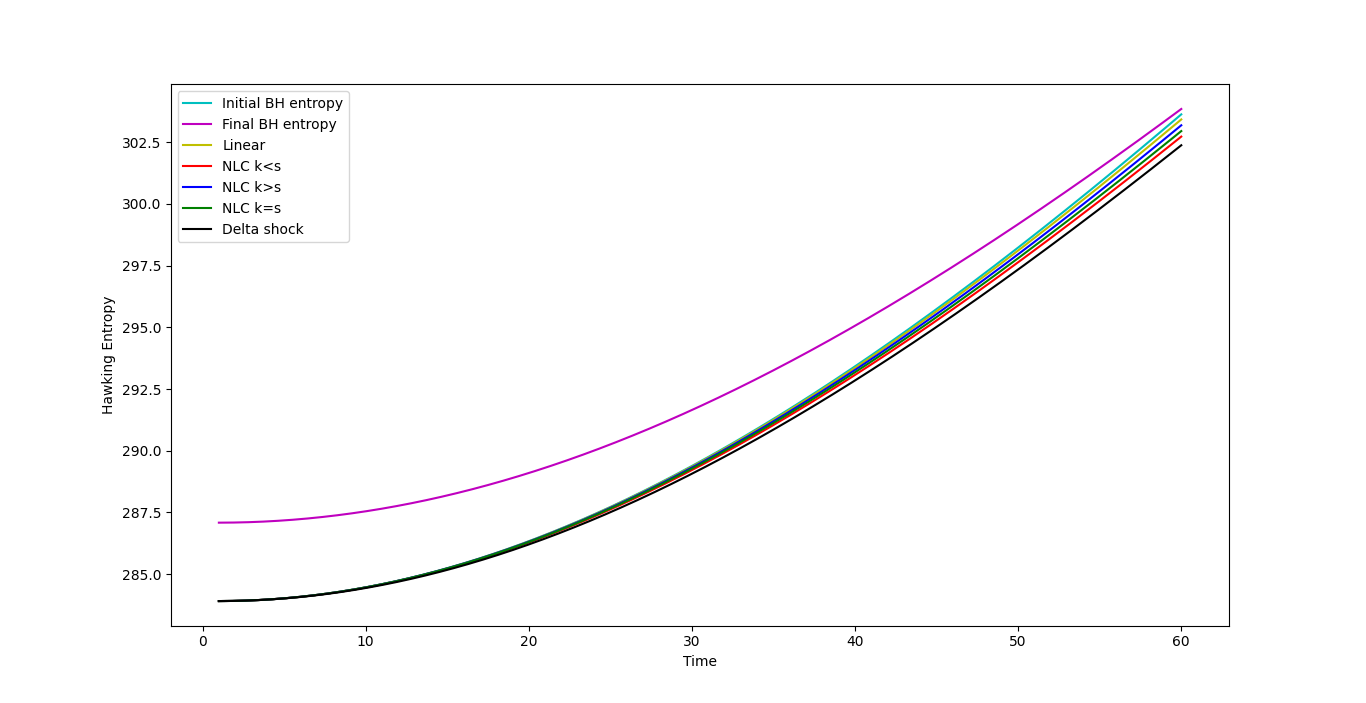}
    \caption{Hawking saddle entropy. \small The different colors represent different bath cooling profiles. Yellow: bath temperature decreasing linearly, Black: a single instant drop in bath temperature, Red/Green/Blue: bath temperature follows Newton's law of cooling (NLC) for $(k < s), (k = s),\text{ and } (k > s)$ cases, respectively.}
    \label{figure10}
\end{figure}

 For Hawking radiation as well, the slopes are changing at different rates and finally converging to the slopes of the entropy curve for the final-state black hole.

\subsection{Linear Decrease in Incoming Stress Tensor}
Linear decay of the incoming stress tensor models scenarios in which the bath energy is drained at a constant rate. This provides a simple toy model that isolates the role of a time-dependent flux.

In this case, we take $\frac{\pi c}{12\tilde{\beta_{t}}^2} = \frac{\pi c}{12\beta^2}-st$. Here, $s(=0.06 \text{chosen for illustrative purposes})$ is the rate of decay in the incoming energy. We follow the same numerical procedure of solving the differential equations to calculate the island entropy for this evaporation profile.

Figures \ref{figure8a}, \ref{figure8b}, \ref{figure10} show the change in energy, the island saddle entropy, and the Hawking saddle entropy over time, respectively.

\subsection{Quasi-static Evaporation}

We next compare the results from the different models of evaporation with the quasi-static case. The quasi-static case is one where the equilibrium is maintained at each time step. The quasi-static process is realized as sequential small drops in bath temperature with long waiting periods between drops to allow the black hole to come to equilibrium with the temperature. This implies that, when evaluated at successive equilibrium configurations of the black hole, the energy balance equation is identically satisfied at each stage. Since we treat the temperature of the bath and the black hole as equal and constant in each time step, the resulting boundary parameterization will be $f(t)=\frac{\beta_t}{\pi}tanh(\frac{\pi}{\beta_t}t)$.

This form of boundary parameterization is obtained precisely when $\{f,t\}=constant$ for each time step. This is precisely what distinguishes this case from the dynamical cases where $\{f,t\}\propto E(t) \neq constant$ at each time step and $f(t)$ had to be computed numerically. Varying $\beta_t$ generates quasi-static evolution. Note that this quasi-static evolution does not represent a dynamical solution of the coupled system, but rather a sequence of equilibrium geometries. We use it as a benchmark to compare with other cooling profiles.

We take $\frac{1}{{\tilde{\beta}^2}_{t}} = \left(\frac{1}{\tilde\beta^2}+\left(\frac{1}{\beta^2}-\frac{1}{\tilde\beta^2}\right)e^{-kt}\right)$.  

\begin{figure}[H]
\begin{subfigure}{0.46\textwidth}
\centering
    \includegraphics[scale=0.4]{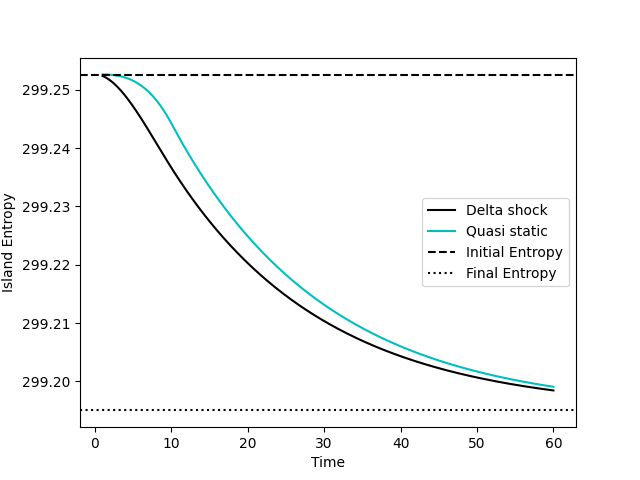}
    \caption{\small Entropy of Island saddle vs Boundary time.}
    \label{figure12a}
\end{subfigure}\quad
\begin{subfigure}{0.46\textwidth}
    \centering
    \includegraphics[scale=0.4]{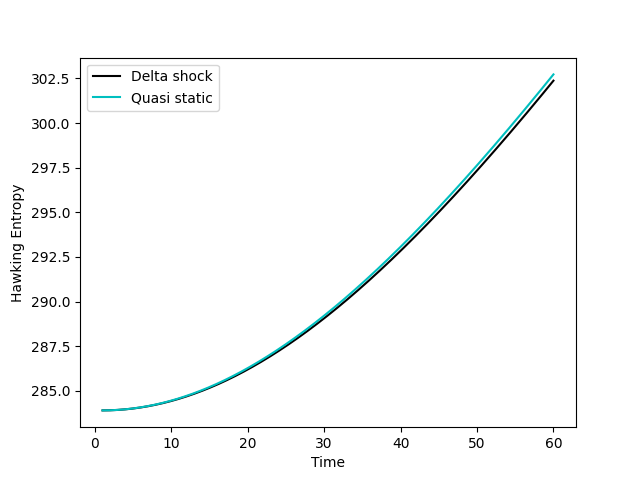}
    \caption{Entropy of Hawking saddle vs Boundary time.}
    \label{fig12}
\end{subfigure}
    \caption{Quasi-static vs Instantaneous (delta-function shock) change in bath temperature.}
\end{figure}

\section{Conclusion}
In this paper, we extended previous results on Page curves for evaporating black holes in JT gravity to more general cases where the bath temperature can be controlled externally. 

We computed the Hawking and the island saddle entropies numerically for an evaporating black hole in an environment with varying temperature. Our approach extends previous discussions of islands in the presence of shockwaves. We continuously change the bath temperature and consider several different profiles of continuous temperature change. In this paper, we studied three profiles: linear change, Newton's law of cooling, and step-function cooling. The result gives the Page curves of a slowly evaporating black hole in AdS under different external conditions of cooling, while remaining in the semiclassical regime.  

Our numerical methods are entirely general and can be used to study any cooling profile of the bath. Our results may provide a useful framework for modeling black hole evaporation under general external conditions. 

An interesting extension of our analysis would be to study stochastic fluctuations in the bath temperature, modeling random or noisy environments. Since our method only requires a time-dependent specification of the incoming stress tensor, incorporating stochasticity is straightforward and would allow one to probe how robust the Page curve is to environmental noise.

Another natural generalization concerns coupling the black hole to multiple baths with competing temperatures. This would allow for the exploration of steady-state non-equilibrium situations where radiation continuously flows through the black hole system. 

We plan to return to these problems in the future.

\section{Acknowledgments}
 NK was supported by the SERB Start-up Research Grant SRG/2022/000970 during a significant part of this research. 

\bibliography{main}

@article{Page:1993df,
    author = "Page, Don N.",
    title = "{Average entropy of a subsystem}",
    eprint = "gr-qc/9305007",
    archivePrefix = "arXiv",
    reportNumber = "ALBERTA-THY-22-93",
    doi = "10.1103/PhysRevLett.71.1291",
    journal = "Phys. Rev. Lett.",
    volume = "71",
    pages = "1291--1294",
    year = "1993"
}

@article{Cadoni:2021ypx,
   title={Unitarity and Page Curve for Evaporation of 2D AdS Black Holes},
   volume={24},
   ISSN={1099-4300},
   url={http://dx.doi.org/10.3390/e24010101},
   DOI={10.3390/e24010101},
   number={1},
   journal={Entropy},
   publisher={MDPI AG},
   author={Cadoni, Mariano and Sanna, Andrea P.},
   year={2022},
   month=Jan, pages={101} }

@article{Almheiri:2019hni,
    author = "Almheiri, Ahmed and Mahajan, Raghu and Maldacena, Juan and Zhao, Ying",
    title = "{The Page curve of Hawking radiation from semiclassical geometry}",
    eprint = "1908.10996",
    archivePrefix = "arXiv",
    primaryClass = "hep-th",
    doi = "10.1007/JHEP03(2020)149",
    journal = "JHEP",
    volume = "03",
    pages = "149",
    year = "2020"
}

@article{Almheiri_2019,
   title={The entropy of bulk quantum fields and the entanglement wedge of an evaporating black hole},
   volume={2019},
   ISSN={1029-8479},
   url={http://dx.doi.org/10.1007/JHEP12(2019)063},
   DOI={10.1007/jhep12(2019)063},
   number={12},
   journal={Journal of High Energy Physics},
   publisher={Springer Science and Business Media LLC},
   author={Almheiri, Ahmed and Engelhardt, Netta and Marolf, Donald and Maxfield, Henry},
   year={2019},
   month=dec }

@article{Penington:2019npb,
    author = "Penington, Geoffrey",
    title = "{Entanglement Wedge Reconstruction and the Information Paradox}",
    eprint = "1905.08255",
    archivePrefix = "arXiv",
    primaryClass = "hep-th",
    doi = "10.1007/JHEP09(2020)002",
    journal = "JHEP",
    volume = "09",
    pages = "002",
    year = "2020"
}

@article{Engelhardt:2014gca,
    author = "Engelhardt, Netta and Wall, Aron C.",
    title = "{Quantum Extremal Surfaces: Holographic Entanglement Entropy beyond the Classical Regime}",
    eprint = "1408.3203",
    archivePrefix = "arXiv",
    primaryClass = "hep-th",
    doi = "10.1007/JHEP01(2015)073",
    journal = "JHEP",
    volume = "01",
    pages = "073",
    year = "2015"
}

@article{Penington:2019kki,
    author = "Penington, Geoff and Shenker, Stephen H. and Stanford, Douglas and Yang, Zhenbin",
    title = "{Replica wormholes and the black hole interior}",
    eprint = "1911.11977",
    archivePrefix = "arXiv",
    primaryClass = "hep-th",
    doi = "10.1007/JHEP03(2022)205",
    journal = "JHEP",
    volume = "03",
    pages = "205",
    year = "2022"
}

@article{Almheiri:2019qdq,
    author = "Almheiri, Ahmed and Hartman, Thomas and Maldacena, Juan and Shaghoulian, Edgar and Tajdini, Amirhossein",
    title = "{Replica Wormholes and the Entropy of Hawking Radiation}",
    eprint = "1911.12333",
    archivePrefix = "arXiv",
    primaryClass = "hep-th",
    doi = "10.1007/JHEP05(2020)013",
    journal = "JHEP",
    volume = "05",
    pages = "013",
    year = "2020"
}

@article{Almheiri:2020cfm,
    author = "Almheiri, Ahmed and Hartman, Thomas and Maldacena, Juan and Shaghoulian, Edgar and Tajdini, Amirhossein",
    title = "{The entropy of Hawking radiation}",
    eprint = "2006.06872",
    archivePrefix = "arXiv",
    primaryClass = "hep-th",
    doi = "10.1103/RevModPhys.93.035002",
    journal = "Rev. Mod. Phys.",
    volume = "93",
    number = "3",
    pages = "035002",
    year = "2021"
}

@misc{Almheiri:2019yqk,
      title={Islands outside the horizon}, 
      author={Ahmed Almheiri and Raghu Mahajan and Juan Maldacena},
      year={2023},
      eprint={1910.11077},
      archivePrefix={arXiv},
      primaryClass={hep-th},
      url={https://arxiv.org/abs/1910.11077}, 
}

@article{Marolf:2020xie,
    author = "Marolf, Donald and Maxfield, Henry",
    title = "{Transcending the ensemble: baby universes, spacetime wormholes, and the order and disorder of black hole information}",
    eprint = "2002.08950",
    archivePrefix = "arXiv",
    primaryClass = "hep-th",
    doi = "10.1007/JHEP08(2020)044",
    journal = "JHEP",
    volume = "08",
    pages = "044",
    year = "2020"
}

@article{Almheiri:2019psy,
    author = "Almheiri, Ahmed and Mahajan, Raghu and Santos, Jorge E.",
    title = "{Entanglement islands in higher dimensions}",
    eprint = "1911.09666",
    archivePrefix = "arXiv",
    primaryClass = "hep-th",
    doi = "10.21468/SciPostPhys.9.1.001",
    journal = "SciPost Phys.",
    volume = "9",
    number = "1",
    pages = "001",
    year = "2020"
}

@article{Chen:2020uac,
    author = "Chen, Hong Zhe and Myers, Robert C. and Neuenfeld, Dominik and Reyes, Ignacio A. and Sandor, Joshua",
    title = "{Quantum Extremal Islands Made Easy, Part I: Entanglement on the Brane}",
    eprint = "2006.04851",
    archivePrefix = "arXiv",
    primaryClass = "hep-th",
    doi = "10.1007/JHEP10(2020)166",
    journal = "JHEP",
    volume = "10",
    pages = "166",
    year = "2020"
}

@article{Hashimoto:2020cas,
    author = "Hashimoto, Koji and Iizuka, Norihiro and Matsuo, Yoshinori",
    title = "{Islands in Schwarzschild black holes}",
    eprint = "2004.05863",
    archivePrefix = "arXiv",
    primaryClass = "hep-th",
    reportNumber = "OU-HET-1053",
    doi = "10.1007/JHEP06(2020)085",
    journal = "JHEP",
    volume = "06",
    pages = "085",
    year = "2020"
}

@article{Hartman:2020khs,
    author = "Hartman, Thomas and Jiang, Yikun and Shaghoulian, Edgar",
    title = "{Islands in cosmology}",
    eprint = "2008.01022",
    archivePrefix = "arXiv",
    primaryClass = "hep-th",
    doi = "10.1007/JHEP11(2020)111",
    journal = "JHEP",
    volume = "11",
    pages = "111",
    year = "2020"
}

@article{Gautason:2020tmk,
    author = "Gautason, Fridrik Freyr and Schneiderbauer, Lukas and Sybesma, Watse and Thorlacius, Larus",
    title = "{Page Curve for an Evaporating Black Hole}",
    eprint = "2004.00598",
    archivePrefix = "arXiv",
    primaryClass = "hep-th",
    doi = "10.1007/JHEP05(2020)091",
    journal = "JHEP",
    volume = "05",
    pages = "091",
    year = "2020"
}

@article{Balasubramanian:2020xqf,
    author = "Balasubramanian, Vijay and Kar, Arjun and Ugajin, Tomonori",
    title = "{Islands in de Sitter space}",
    eprint = "2008.05275",
    archivePrefix = "arXiv",
    primaryClass = "hep-th",
    doi = "10.1007/JHEP02(2021)072",
    journal = "JHEP",
    volume = "02",
    pages = "072",
    year = "2021"
}

@article{Krishnan:2020fer,
    author = "Krishnan, Chethan",
    title = "{Critical Islands}",
    eprint = "2007.06551",
    archivePrefix = "arXiv",
    primaryClass = "hep-th",
    doi = "10.1007/JHEP01(2021)179",
    journal = "JHEP",
    volume = "01",
    pages = "179",
    year = "2021"
}

@misc{Krishnan:2020oun,
      title={Page Curve and the Information Paradox in Flat Space}, 
      author={Chethan Krishnan and Vaishnavi Patil and Jude Pereira},
      year={2020},
      eprint={2005.02993},
      archivePrefix={arXiv},
      primaryClass={hep-th},
      url={https://arxiv.org/abs/2005.02993}, 
}

@article{Ling:2020laa,
    author = "Ling, Yi and Liu, Yuxuan and Xian, Zhuo-Yu",
    title = "{Island in Charged Black Holes}",
    eprint = "2010.00037",
    archivePrefix = "arXiv",
    primaryClass = "hep-th",
    doi = "10.1007/JHEP03(2021)251",
    journal = "JHEP",
    volume = "03",
    pages = "251",
    year = "2021"
}

@article{Ahn:2021chg,
    author = "Ahn, Byoungjoon and Bak, Sang-Eon and Jeong, Hyun-Sik and Kim, Keun-Young and Sun, Ya-Wen",
    title = "{Islands in charged linear dilaton black holes}",
    eprint = "2107.07444",
    archivePrefix = "arXiv",
    primaryClass = "hep-th",
    doi = "10.1103/PhysRevD.105.046012",
    journal = "Phys. Rev. D",
    volume = "105",
    number = "4",
    pages = "046012",
    year = "2022"
}

@article{Hollowood:2020cou,
    author = "Hollowood, Timothy J. and Kumar, S. Prem",
    title = "{Islands and Page Curves for Evaporating Black Holes in JT Gravity}",
    eprint = "2004.14944",
    archivePrefix = "arXiv",
    primaryClass = "hep-th",
    doi = "10.1007/JHEP08(2020)094",
    journal = "JHEP",
    volume = "08",
    pages = "094",
    year = "2020"
}

@article{Jackiw:1984je,
    author = "Jackiw, R.",
    editor = "Baier, R. and Satz, H.",
    title = "{Lower Dimensional Gravity}",
    reportNumber = "MIT-CTP-1203",
    doi = "10.1016/0550-3213(85)90448-1",
    journal = "Nucl. Phys. B",
    volume = "252",
    pages = "343--356",
    year = "1985"
}

@article{Teitelboim:1983ux,
    author = "Teitelboim, C.",
    title = "{Gravitation and Hamiltonian Structure in Two Space-Time Dimensions}",
    doi = "10.1016/0370-2693(83)90012-6",
    journal = "Phys. Lett. B",
    volume = "126",
    pages = "41--45",
    year = "1983"
}

@article{Maldacena:2016hyu,
    author = "Maldacena, Juan and Stanford, Douglas",
    title = "{Remarks on the Sachdev-Ye-Kitaev model}",
    eprint = "1604.07818",
    archivePrefix = "arXiv",
    primaryClass = "hep-th",
    doi = "10.1103/PhysRevD.94.106002",
    journal = "Phys. Rev. D",
    volume = "94",
    number = "10",
    pages = "106002",
    year = "2016"
}

@article{Almheiri:2014cka,
    author = "Almheiri, Ahmed and Polchinski, Joseph",
    title = "{Models of AdS$_{2}$ backreaction and holography}",
    eprint = "1402.6334",
    archivePrefix = "arXiv",
    primaryClass = "hep-th",
    doi = "10.1007/JHEP11(2015)014",
    journal = "JHEP",
    volume = "11",
    pages = "014",
    year = "2015"
}

@article{Jensen:2016pah,
    author = "Jensen, Kristan",
    title = "{Chaos in AdS$_2$ Holography}",
    eprint = "1605.06098",
    archivePrefix = "arXiv",
    primaryClass = "hep-th",
    doi = "10.1103/PhysRevLett.117.111601",
    journal = "Phys. Rev. Lett.",
    volume = "117",
    number = "11",
    pages = "111601",
    year = "2016"
}

@article{Maldacena:2016upp,
    author = "Maldacena, Juan and Stanford, Douglas and Yang, Zhenbin",
    title = "{Conformal symmetry and its breaking in two dimensional Nearly Anti-de-Sitter space}",
    eprint = "1606.01857",
    archivePrefix = "arXiv",
    primaryClass = "hep-th",
    doi = "10.1093/ptep/ptw124",
    journal = "PTEP",
    volume = "2016",
    number = "12",
    pages = "12C104",
    year = "2016"
}

@article{Engelsoy:2016xyb,
    author = {Engels{\"o}y, Julius and Mertens, Thomas G. and Verlinde, Herman},
    title = "{An investigation of AdS$_{2}$ backreaction and holography}",
    eprint = "1606.03438",
    archivePrefix = "arXiv",
    primaryClass = "hep-th",
    doi = "10.1007/JHEP07(2016)139",
    journal = "JHEP",
    volume = "07",
    pages = "139",
    year = "2016"
}

@article{Mertens:2022irh,
    author = "Mertens, Thomas G. and Turiaci, Gustavo J.",
    title = "{Solvable models of quantum black holes: a review on Jackiw{\textendash}Teitelboim gravity}",
    eprint = "2210.10846",
    archivePrefix = "arXiv",
    primaryClass = "hep-th",
    doi = "10.1007/s41114-023-00046-1",
    journal = "Living Rev. Rel.",
    volume = "26",
    number = "1",
    pages = "4",
    year = "2023"
}

@article{Mertens:2019bvy,
    author = "Mertens, Thomas G.",
    title = "{Towards Black Hole Evaporation in Jackiw-Teitelboim Gravity}",
    eprint = "1903.10485",
    archivePrefix = "arXiv",
    primaryClass = "hep-th",
    doi = "10.1007/JHEP07(2019)097",
    journal = "JHEP",
    volume = "07",
    pages = "097",
    year = "2019"
}

@article{Calabrese:2004eu,
    author = "Calabrese, Pasquale and Cardy, John L.",
    title = "{Entanglement entropy and quantum field theory}",
    eprint = "hep-th/0405152",
    archivePrefix = "arXiv",
    doi = "10.1088/1742-5468/2004/06/P06002",
    journal = "J. Stat. Mech.",
    volume = "0406",
    pages = "P06002",
    year = "2004"
}
\end{document}